\documentclass[singlecolumn,epjc3]{svjour3}
\smartqed  
\RequirePackage{graphicx}
\journalname{Eur. Phys. J. C}

\begin{document}

\title{Modeling Anisotropic Stars Obeying Chaplygin Equation of State}

\author{Piyali Bhar\thanksref{e1,addr1} \and Megan Govender\thanksref{e2,addr2} \and Ranjan Sharma\thanksref{e3,addr3}}
\thankstext{e1}{e-mail:piyalibhar90@gmail.com }
\thankstext{e2}{e-mail:megandhreng@dut.ac.za}
\thankstext{e3}{e-mail:rsharma@associates.iucaa.in}

\institute{Department of Mathematics,Government General Degree College, Singur, Hooghly 712 409, West Bengal,
India\label{addr1}
\and
Department of Mathematics, Faculty of Applied Sciences, Durban University of Technology, Durban, 4000, South Africa\label{addr2}
\and
Department of Physics, P. D. Women's College, Jalpaiguri 735101, India.\label{addr3}}

\date{Received: date / Accepted: date}

\maketitle

\begin{abstract}
In this work we provide a framework for modeling compact stars in which the interior matter distribution obeys a generalised Chaplygin equation of state.
The interior geometry of the stellar object is described by a spherically symmetric line element which is simultaneously comoving and isotropic with the exterior spacetime being vacuum. We are able to integrate the Einstein field equations and present closed form solutions which adequately describe compact strange star candidates like Her X-1, RX J 1856-37, PSRJ 1614-2230 and SAX J1808.4-3658.
\end{abstract}

\keywords{General Relativity, Anisotropy, Compact star, Chaplygin equation of state}

\maketitle

\section{Introduction}
The search for exact solutions of the Einstein field equations has generated a rich field of models describing relativistic compact objects. Since the pioneering work of Schwarzschild who obtained the first interior solution describing a uniform density sphere, the modeling of relativistic stars has moved from the regime of toy models to sophisticated, realistic stellar structures. With the discovery of pulsars, neutron stars and strange stars there was a need to obtain relativistic analogues of Newtonian stars, particularly when the densities of the stellar material was of the order of $10^{14}$gm~cm$^{-3}$. The simplistic model of a static uniform density star has been generalised to include the effects of pressure anisotropy, electric charge, scalar field, dark energy and the cosmological constant, on the gross physical properties of compact objects\cite{rayso,ma1,ma2}. Models of relativistic fluid spheres have also been obtained within the framework of higher order theories of gravity including the Randall-Sundrum brane scenario, Einstein-Gauss-Bonnet gravity and Lovelock formalisms\cite{sud1,sud2,abbas,nare}. In order to close the system of equations governing the gravitational and thermodynamical behaviour of bounded objects, various techniques were employed by researchers working in this field of study: (i) imposition of symmetry, (ii) adhoc assumptions of the gravitational potentials, (iii) specific choices of the fall-off behaviour of the pressure, density or the anisotropy, to name a few\cite{dev1,dev2,iv1}. To construct a stellar model, a physically motivated route, in general, is to impose an equation of state which relates the pressure as a thermodynamical function of the density, ie., $p = p(\rho)$. Most of the earlier works were centered on imposing a linear equation of state of the form $p = \alpha \rho$ where $\alpha$ is a constant. This was later generalised to $p = \alpha \rho - \beta$, where $\beta = \alpha \rho_s$ and $\rho_s$ is the surface density\cite{sm1}. The conditions were relaxed by allowing for anisotropic pressure. Note that works in fundamental particle physics led to the MIT-bag model which hinged on an equation of state of the form $p = \alpha \rho - 4B$ where $B$ is the Bag constant. The linear equation of state was further generalised to the quadratic equation of state of the form $p = \alpha \rho - \beta + \sigma \rho^2$\cite{sifiso}. One of the first successful attempts to obtain a generalisation of the Newtonian polytrope was achieved by Buchdahl\cite{buch} in which he obtained a pseudo-relativistic version of the Lane-Emden polytrope of index $5$. Herrera and Barreto presented a general formalism to generate relativistic polytropes with anisotropic pressure in Schwarzschild coordinates\cite{poly}. Their findings also prompted further investigations into the origins of anisotropy, cracking in relativistic stellar models and stability\cite{azam}. In order to fine-tune these models with observations, some researchers employed a mixed polytrope equation of state in which two or more species of particles made up the stellar fluid. The inclusion of charge within the stellar core led to a plethora of static stellar models in which the role of the electromagnetic field on the stability, mass-radius ratio and redshift was demonstrated\cite{r1,ra1,ra2}.

Issues surrounding the black hole horizon paradox necessitated the search for alternative models of black holes free of horizons.
The gravastar model was first proposed by Mazur and Mottola \cite{mazur} which sought to address many of the problems encountered during the final stages of gravitational collapse. Dynamically the model hinged on the phenomenon that during the latter epoch of gravitational collapse spacetime itself would undergo phase transitions which would halt collapse.
The emerging picture of a gravastar was that of a layered composite:
a de Sitter interior filled with constant positive (dark)
 energy density $\rho$ featuring an isotropic negative pressure $\rho = -p >0$. This layer is then connected
via three intermediate layers to an exterior vacuum Schwarzschild solution.  The intermediate relatively thin shell is composed of stiff matter ($p = \rho$). Stability of the composite profile is achieved by utilising two infinitesimally-thin shells endowed with surface densities $\sigma_\pm$ and surface tensions $\vartheta_\pm$.  An interesting model was proposed by Usmani {\em et al}\cite{ra2} in which they generalised the Mazur-Mottola gravastar picture to include charge. In addition, the interior of the gravastar admitted conformal motion. They were able to show that charged interior de Sitter
void must generate the gravitational mass. This mass is accountable for the attractive force that counter-balances the electromagnetic repulsion due to the presence of charge during the collapse process\cite{rahaman1}.  In a more recent model, Banerjee {\em et al}\cite{brig} presented a Braneworld generalisation of a gravastar admitting conformal motion. Motivated by the existence of dark energy, Lobo and coworkers\cite{lobo1} have proposed stellar models, the so-called `dark stars' in which the equation of state of is of the form $p = \alpha \rho$ in which $-1 < \alpha < -1/3$. It has been proposed that in the phantom regime ($\alpha = -1$), the extremely high pressures may invoke a topological change rendering the dark energy star to a wormhole. An interesting proposal regarding dark energy and dark matter is treating them as  different manifestations of a single entity. This proposal leads to the Chaplygin gas model in which the equation of state derives from string theory. Various applications of the Chaplygin gas  model have been pursued in order to account for cosmological observations such as acceleration of the cosmic fluid and structure formation. The Chaplygin equation of state has been subsequently modified to a more generalized Chaplygin gas equation of state. The generalised Chaplygin equation of state has been employed to model dark stars which are remnants of continued gravitational collapse. The idea here is that the dark energy provides sufficient repulsion to halt collapse leading to stable bounded configurations free of horizons and singularities\cite{p1,p2}.

This paper is structured as follows: In section $2$ we introduce the field equations necessary for the modeling of a spherically symmetric star within the framework of general relativity. In section $3$. we present a particular solution describing the interior of the star in which the matter content obeys a generalised Chaplygin equation of state. The junction conditions required for the smooth matching of the interior spacetime to the exterior Schwarzschild solution are worked out in section $4$. A detailed physical analysis of the geometrical and thermodynamical behaviour of our model is presented in section $5$. We discuss the stability, energy conditions and mass-radius relation in sections $6$, $7$ and $8$, respectively. We conclude with a discussion of our results in section $9$.

\section{Spherically symmetric spacetime}

We consider a model which represents a static spherically symmetric anisotropic fluid configuration obeying a generalised Chaplygin
equation of state. The interior spacetime is described by a spherically symmetric line element which is simultaneously comoving and isotropic
\begin{equation}
ds^2 = -A^2(r)dt^2 + B^2(r)\left[dr^2 + r^2d\Omega^2\right],
\label{metric}
\end{equation}
where $d\Omega^2 = d\theta^2 + \sin^2{\theta}d\phi^2$ and the
metric functions, $A(r)$ and $B(r)$ are yet to be determined.
 For
our model the energy-momentum tensor for the stellar fluid is
\begin{equation} T_{ab} = {\mbox diag}\left(-\rho, p_r, p_t, p_t\right),\label{2}
\end{equation} where $\rho$, $p_r$ and $p_t$ are the proper energy density,
radial pressure and tangential pressure, respectively. The fluid
four--velocity ${\bf u}$ is comoving and is given by
\begin{equation} u^a = \displaystyle\frac{1}{A} \delta^{a}_0 \,.
\label{2'}
\end{equation}
The Einstein field equations for the line element (\ref{metric})
are
\begin{equation}\label{g3a}
8\pi\rho = - \frac{1}{B^2} \left(
2\frac{B''}{B} - \frac{{B'}^2}{B^2} +
\frac{4}{r}\frac{B'}{B} \right),
\end{equation}
\begin{equation}\label{g3b}
8\pi p_r = \frac{1}{B^2} \left[\frac{{B'}^2}{B^2} +
2\frac{A'}{A}\frac{B'}{B} + \frac{2}{r} \left(\frac{A'}{A} +
\frac{B'}{B}\right) \right],
\end{equation}
\begin{equation}\label{g3c}
8\pi p_t=\frac{1}{B^2} \left[\frac{A''}{A} + \frac{B''}{B}-
\frac{{B'}^2}{B^2}+\frac{1}{r}
\left(\frac{A'}{A}+\frac{B'}{B}\right)\right],
\end{equation}
where primes denote differentiation with respect to the radial
coordinate $r$. We have utilized geometrized units in deriving the above system of equations in which the coupling constant and the
speed of light are taken to be unity.  The mass of the spherical object is given by
\begin{equation}
\label{mass} m(r)= 4\pi\int_0^r\omega^2
\rho(\omega)d\omega,
\end{equation}
where $\omega$ is an integration variable. In order to close the system of equations, we assume that the interior matter distribution obeys a generalised Chaplygin equation of state of the form
\begin{equation} \label{eos}
p_r = H\rho - \frac{K}{\rho}
\end{equation}
where $H$ and $K$ are positive constants. Substituting (\ref{g3a}) and (\ref{g3b}) in (\ref{eos}) we obtain
\begin{equation} \label{master}
\frac{A'}{A} = \frac{1}{2}\left(\frac{B'}{B}+\frac{1}{r}\right)^{-1}[G(r)-F(r)],
\end{equation}
where \begin{equation}
G(r) =\frac{(8\pi)^{2}KB^{4}}{ 2\frac{B''}{B} - \frac{{B'}^2}{B^2} +
\frac{4}{r}\frac{B'}{B}},\end{equation}
\begin{equation}
F(r)=2H\frac{B''}{B}+(1-H)\left(\frac{B'}{B}\right)^{2}+(1+2H)\frac{2}{r}\frac{B'}{B}.
\end{equation}

On integrating (\ref{master}) we obtain
\begin{equation}\label{g5}
A= d \exp \left[\frac{1}{2}\int H(r) dr\right],
\end{equation}
where
\begin{equation} \label{h} H(r) =  \left(\frac{B'}{B}+\frac{1}{r}\right)^{-1}[G(r)-F(r)],
\end{equation}
and $d$ is a constant of integration. Therefore, the line element
(\ref{metric}) can now be written as
\begin{equation}\label{g6}
ds^2 = -d^2 \exp \left[\int H(r) dr\right] dt^2 + B^2 \left[dr^2
+ r^2d\Omega^2\right],
\end{equation}
where $H(r)$ is given in (\ref{h}). Hence, any solution describing
a static spherically symmetric anisotropic matter distribution
obeying a generalised Chaplygin equation of state  in isotropic coordinates can be
easily determined by a single generating function $B(r)$.

\section{Generating solutions}
In order to close the system of equations several choices for $B(r)$ can be made. It is interesting to note that the choice of the metric potential $B(r)$ determines the gravitational and thermodynamical behaviour of the model. Hence the choice of $B(r)$ must satisfy all the requirements for a realistic stellar model. Recent work by Naidu and Govender\cite{ng2016} have shown that the end-state of gravitational collapse resulting from a dynamically unstable static core is `sensitive' to the choice of the initial metric functions. They show that for the same $B(r)$ but with two distinct initially static cores; (i) vanishing radial pressure within the static configuration and (ii) uniform density interior, the final outcome of dissipative collapse leads to very different temperature profiles. Following Govender and Thirukkanesh\cite{gov1} we utilise the physically motivated choice for $B(r)$ as
\begin{equation}\label{b}
B(r) = \frac{a}{\sqrt{1 + br^2}}
\end{equation}
where $a$ and $b$ are constants. One can easily verify that the gravitational potential
$B$ in (\ref{b}) satisfies the regularity conditions, $B(0)$ = constant and $B'(r) = 1$ at the
origin. The same expression of $B(r)$ was previously utilized to model compact objects in curvature coordinates by Schwarzschild \cite{sch}, Einstein \cite{ein} and de
Sitter \cite{des} and more recently in  comoving coordinates by Govender and Thirukkanesh \cite{gov1} and Thirukkanesh {\em et al.} \cite{thir}.\\
With this choice of $B(r)$, we obtain from (\ref{g5})
\[
A(r) = d\exp\left[\left\{\frac{1+H}{4}-\frac{16a^{4}K\pi^{2}}{b^{2}}\right\}(1+br^{2})\right](1 + br^2)^{\frac{1+5H}{4}}(6 + br^2)^{\frac{80a^4K\pi^{2}}{b^2}},\]
where $d$ is a constant of integration.\\
Subsequently, the field equations yield
\begin{eqnarray}
\rho &=& \frac{b(6 + br^2)}{8\pi a^2(1 + br^2)},  \\
p_r &=& \frac{bH}{8\pi a^2}\left(\frac{6 + br^2}{1 + br^2}\right) - \frac{8\pi a^2K}{b}\left(\frac{1 + br^2}{6 + br^2}\right),\\
p_t &=& \frac{C_1+C_2r^2+C_3r^4+C_4r^6+C_5r^8
C_6r^{10}}{32\pi a^2 b^2 (6 + b r^2)^2 (1 + b r^2)},
\end{eqnarray}
where, $C_i's$ (i=1,2,...6) are given by
\[C_1=96 b (9 b^2 H - 16 a^4 K \pi^2)\]
\[C_2=16 [9 b^4 \{3 + H (10 + 9 H)\} - 96 a^4 b^2 (5 + 3 H) K \pi^2 +
   256 a^8 K^2 \pi^4]\]
\[C_3=8 b [3 b^4 \{18 + H (47 + 36 H)\} - 32 a^4 b^2 (47 + 42 H) K \pi^2 +
   2048 a^8 K^2 \pi^4]\]
   \[C_4=8 b^2 [b^4 \{18 + H (43 + 27 H)\} - 16 a^4 b^2 (57 + 61 H) K \pi^2 +
   3072 a^8 K^2 \pi^4]\]
\[C_5=4 b^3 \{b^2 (1 + H) - 64 a^4 K \pi^2\} \{b^2 (5 + 6 H) -
   64 a^4 K \pi^2\}\]
\[C_6=b^4 \{b^2 (1 + H) - 64 a^4 K \pi^2\}^2.\]

We define the anisotropic factor as
\begin{equation}
\Delta=p_t-p_r
\end{equation}
which is repulsive in nature if $\Delta>0$ and attractive if $\Delta<0$.

\section{Matching Conditions}
In this section we match the interior spacetime
$({\cal M}_{-})$  to the exterior spacetime $({\cal M}_{+})$
described by the exterior Schwarzschild solution in comoving
isotropic coordinates\cite{bonnor1}
\begin{eqnarray} ds^2 &=& - \frac{\left( 1 -
\frac{M}{2r} \right)^2}{\left( 1 + \frac{M}{2r} \right)^2} dt^2 +
\nonumber\\
&& \left( 1 + \frac{M}{2r} \right)^{4} [dr^2 + r^2 (d\theta^2
+\sin^2\theta d\phi^2)], \label{exterior-metric}
\end{eqnarray}
where $M$ is the mass within a sphere of radius $R$.
Matching  of  interior metric (\ref{metric}) and exterior metric
(\ref{exterior-metric}) at the boundary $r=R$ leads to the
constraints
\begin{eqnarray}
\label{g12} A(R) &=&\frac{\left(1 -\frac{M}{2R}\right)}{\left(1 +\frac{M}{2R}\right)},\\
\label{g13}B(R) &=&\left(1 +\frac{M}{2R}\right)^2,
\end{eqnarray}
where
\begin{equation} \label{masss}
M=m(R)=\frac{bR^3+15\left(R- \frac{\arctan
[\sqrt{b}R]}{\sqrt{b}}\right)}{6a^2}.
\end{equation}
The condition (\ref{g12}) imposes the following restriction on the
constant of integration
\begin{eqnarray}
d&=&\frac{12a^2R -\left[bR^3 +15\left(R- \frac{\arctan
[\sqrt{b}R]}{\sqrt{b}}\right)\right]}{12a^2R +\left[bR^3
+15\left(R- \frac{\arctan
[\sqrt{b}R]}{\sqrt{b}}\right)\right]}  \nonumber\\
&&\times\exp\left[\left\{\frac{16a^{4}K\pi^{2}}{b^{2}}-\frac{1+H}{4}\right\}(1+bR^{2})\right](1 + bR^2)^{-\frac{1+5H}{4}}(6 + bR^2)^{-\frac{80a^4K\pi^{2}}{b^2}}\nonumber\\
\end{eqnarray}
The condition (\ref{g13}) implies
\begin{equation}
\frac{a}{\sqrt{1+b R^2}}= \left[1+\frac{bR^3+15\left(R-
\frac{\arctan [\sqrt{b}R]}{\sqrt{b}}\right)}{12a^2R}\right]^2,
\end{equation}
which imposes a restriction on the parameters $a$ and $b$ which can be determined if we specify the radius of the sphere.

To examine the behabiour of the model parameters like matter density, radial and transverse pressure etc. we assume  $a = 1.3997$, $b=0.009$, and $H=0.294$. By using the matching conditions together with $p_r (r=R) = 0$, we obtain the constant $K = 2.04558 \times 10^{-7}$ for a star of radius $6.7~$km. The mass of the stellar configuration turns out to be $0.789~ M_{\odot}$ which is very close to the observed mass of the  strange star candidate Her X-1\cite{rawls}.

\section{Physical Analysis}
We are now in a position to discuss the the physical features of
the model generated in the preceding section. In order to describe a
realistic stellar structure our model must satisfy the following
physical requirements :

\begin{enumerate}
\item Regularity of the gravitational potentials at the origin:

In our model, $A^2(0)=d^2 6^{\frac{160a^4K \pi^{2}}{b^2}}e^{[b^2(1+{H})-64a^4K\pi^{2}]/2b^2},~ B^2(0)=a^2$
which are constants  and $(A^2(r))'=(B^2(r))'=0 $ at the origin
$r=0$, which indicates that the gravitational potentials are regular at the origin.

\item Positive definiteness of the energy density and pressure at the centre:

Since $\rho(0)= \frac{3b}{4\pi a^2}$, the energy density is positive and regular at the origin. We also have $p_r(0)=\frac{9b^2H - 16\pi^{2}a^4K}{12\pi a^2b}$. To ensure that the radial pressure is positive at the center we must have $\frac{K}{H} < \left(\frac{3b}{4\pi a^2}\right)^{2}$.

Moreover, $\displaystyle
\frac{d\rho}{dr} = -\frac{5b^2r}{4\pi a^2(1+br^2)^2} <0$ i.e., the energy
density is a decreasing function of $r$.

We also note that $\displaystyle\frac{dp_r}{dr} = -\frac{80a^2K \pi r}{(6 + br^2)^2} - \frac{5b^2Hr}{4\pi a^{2}(1 + br^2)^2}$, which implies that $p_r$ is a decreasing function of $r$.

\item Continuity of the extrinsic curvature across the matching hyper-surface, $K^{-}_{ij}
= K^{+}_{ij}$:

Continuity of the extrinsic curvature across the
matching hyper-surface, $r=R$ yields
\begin{equation}
(p_r)_{(r=R)} = 0;\end{equation}
which gives
\[R=\sqrt{\frac{2(32 a^4 K \pi^2+20 a^2 b \pi \sqrt{HK}-3b^{2}H)}{b(b^2 H - 64 a^4 K \pi^2)}},\]
which is finite for appropriate choice of parameters $a, b, H$ and $K$.

\item Ratio of trace of stress tensor to energy density $(p_r+2p_t)/\rho$:

Fulfillment of the requirement that the ratio of trace of stress tensor to energy density $(p_r+2p_t)/\rho$ should decrease radially
outward is shown graphically in Fig.~(\ref{diag}).

\item  Velocity of sound:

For causality to be obeyed the radial and transverse velocities of sound should be in between [$0,~1$].
The radial velocity $(v_{sr}^{2})$ and transverse velocity $(v_{st}^{2})$ of sound can be obtained as
\begin{equation}
v_{sr}^{2}=\frac{dp_r}{d\rho},
\end{equation}
\begin{equation}
v_{st}^{2}=\frac{dp_t}{d\rho}.
\end{equation}
Due to the complexity, we illustrate the causality conditions with the help of graphical representations. Fig.~(\ref{sv1}) and Fig.~(\ref{sv2})clearly show that $0 <v_{sr}^{2}\leq 1$ and $0<v_{st}^{2} \leq 1$ everywhere within the stellar configuration.

\item Stability:

Following Heintzmann and Hillebrandt\cite{hein}, a model of anisotropic compact star will be stable if $\Gamma>\frac{4}{3}$ everywhere within the stellar interior where the adiabatic index $\Gamma$ is defined as
\begin{equation}
\Gamma=\frac{\rho+p_r}{p_r}\frac{dp_r}{d\rho}.
\end{equation}
Fig.~(\ref{gamma}) shows that values of the adiabatic index which clearly indicates that the particular configuration developed in this paper is stable.

\item Energy conditions:

A realistic star should satisfy the energy conditions namely, the Weak Energy Condition (WEC), Null Energy Condition (NEC) and Strong Energy Condition (SEC) as given below:
\begin{equation}\label{ec1}
(i)\,\,\mbox{Null energy condition (NEC)}:  \rho \geq 0,
\end{equation}
\begin{equation}\label{ec2}
(ii) \,\,\mbox{Weak energy condition (WEC)}: \rho-p_r \geq 0, \rho-p_t \geq 0,
\end{equation}
\begin{equation}\label{ec3}
(iii)\,\,\mbox{Strong energy condition (SEC)}: \rho-p_r-2p_t \geq 0.
\end{equation}
For the specific stellar configuration developed here, validity of the inequalities (\ref{ec1})-(\ref{ec3}) have been shown with the help of graphical representations in Fig.~(\ref{ec}).
\end{enumerate}

\begin{figure}[htbp]
   \centering
       \includegraphics[scale=.7]{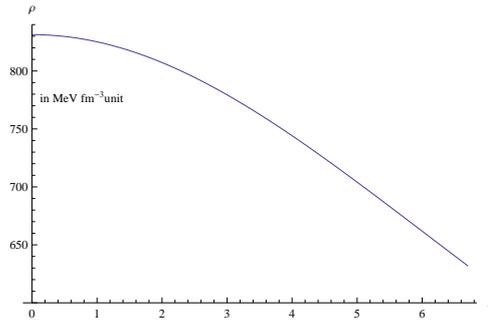}
      \caption{Matter density $\rho$ is plotted against the radial distance $r$ inside the fluid for a particular configuration with $a=1.3997$, $b=0.009$, $H=0.294$ and $K=2.04558\times10^{-7}$}.\label{rho}
\end{figure}

\begin{figure}[htbp]
   \centering
       \includegraphics[scale=.7]{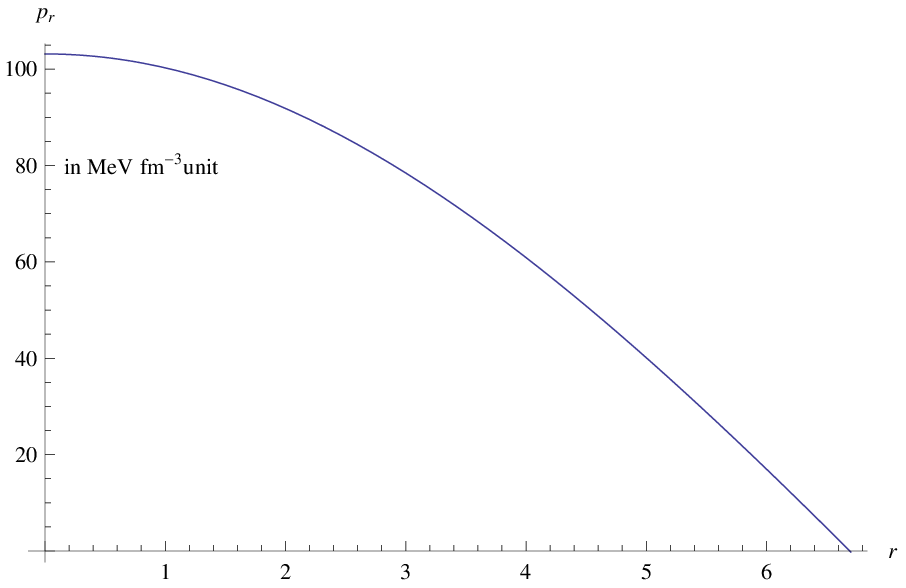}
      \caption{Radial pressure $p_r$ is plotted against the radial distance $r$ inside the stellar interior by employing the same values of the constants as mentioned in Fig.~(\ref{rho}).}\label{pr}
\end{figure}

\begin{figure}[htbp]
   \centering
       \includegraphics[scale=.7]{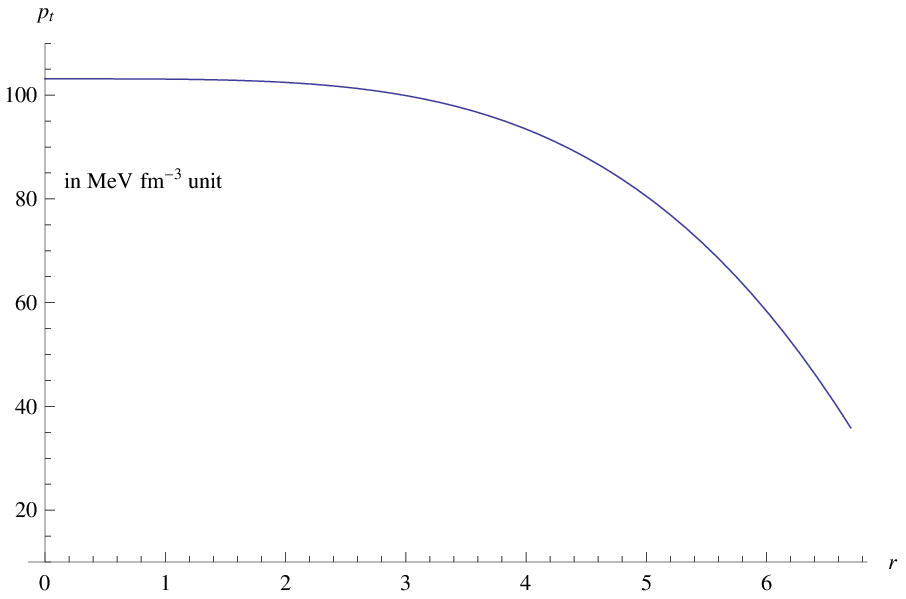}
      \caption{Transverse pressure $p_t$ is plotted against the radial distance $r$ inside the stellar interior by employing the same values of the constants as mentioned in Fig.~(\ref{rho}).}\label{pt}
\end{figure}

\begin{figure}[htbp]
   \centering
       \includegraphics[scale=.7]{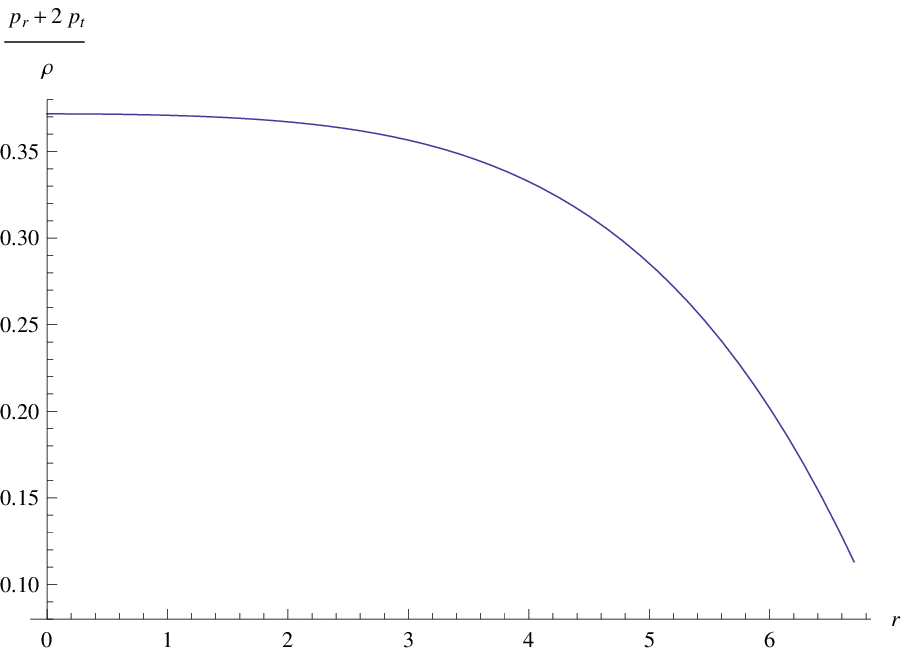}
      \caption{$\frac{p_r+2p_t}{rho}$ is plotted against the radial distance $r$ inside the stellar interior by employing the same values of the constants as mentioned in  Fig.~(\ref{rho}).}\label{diag}
\end{figure}

\begin{figure}[htbp]
   \centering
       \includegraphics[scale=.7]{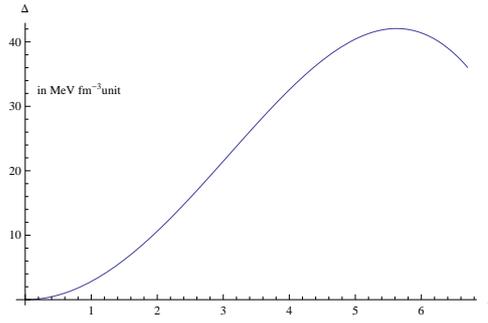}
      \caption{Anisotropic factor is plotted against the radial distance $r$ inside the stellar interior by employing the same values of the constants as mentioned in Fig.~(\ref{rho}).}\label{delta}
\end{figure}

\begin{figure}[htbp]
   \centering
       \includegraphics[scale=.7]{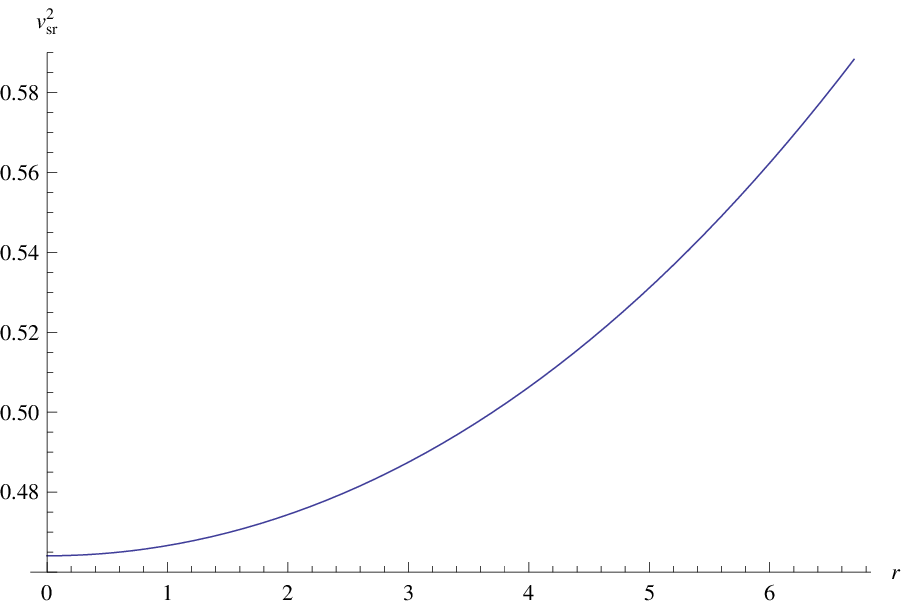}
      \caption{Square of radial velocity of sound is plotted against the radial distance $r$ inside the stellar interior by employing the same values of the constants as mentioned in Fig.~(\ref{rho}).}
   \label{sv1}
\end{figure}

\begin{figure}[htbp]
   \centering
       \includegraphics[scale=.7]{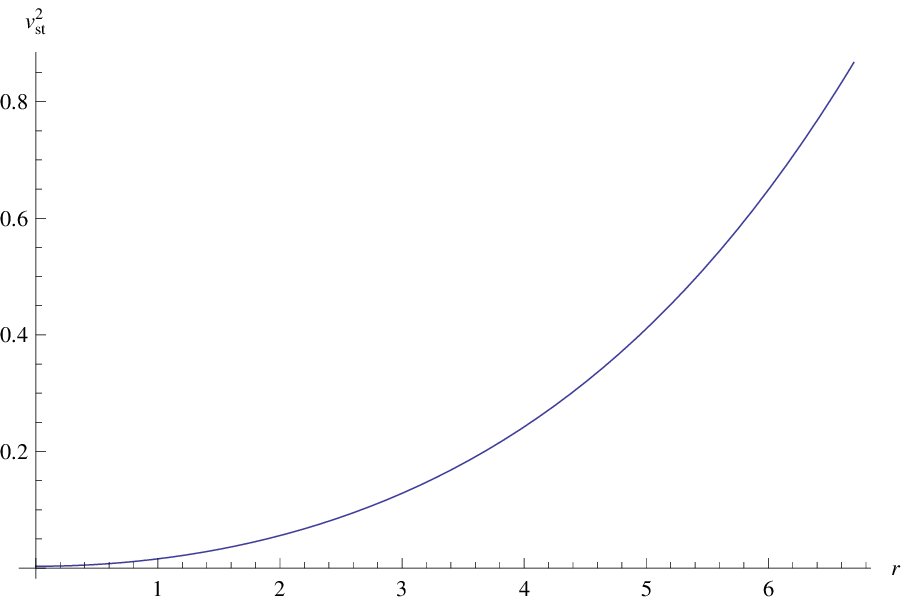}
      \caption{Square of transverse velocity of sound is plotted against the radial distance $r$ inside the stellar interior by employing the same values of the constants as mentioned in Fig.~(\ref{rho}).}
   \label{sv2}
\end{figure}

\begin{figure}[htbp]
   \centering
       \includegraphics[scale=.7]{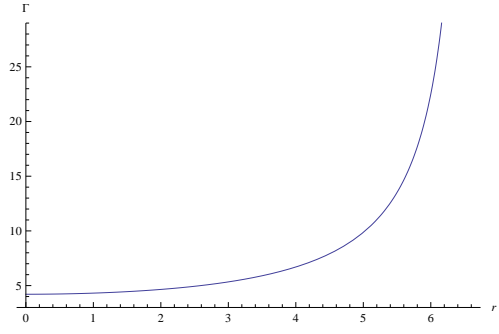}
      \caption{The adiabatic index $\Gamma$ is plotted against the radial distance $r$ inside the stellar interior by employing the same values of the constants as mentioned in Fig.~(\ref{rho}).}
   \label{gamma}
\end{figure}

\begin{figure}[htbp]
   \centering
       \includegraphics[scale=.7]{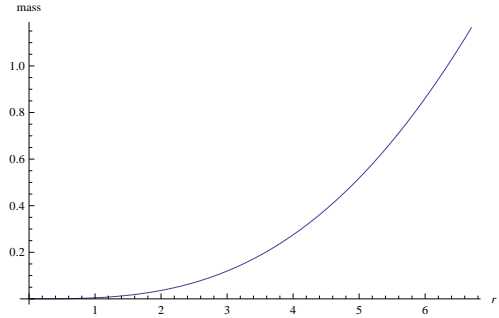}
      \caption{Mass function is is plotted against the radial distance $r$ inside the stellar interior by employing the same values of the constants as mentioned in Fig.~(\ref{rho}).}
   \label{mass}
\end{figure}

\begin{figure}[htbp]
   \centering
       \includegraphics[scale=.7]{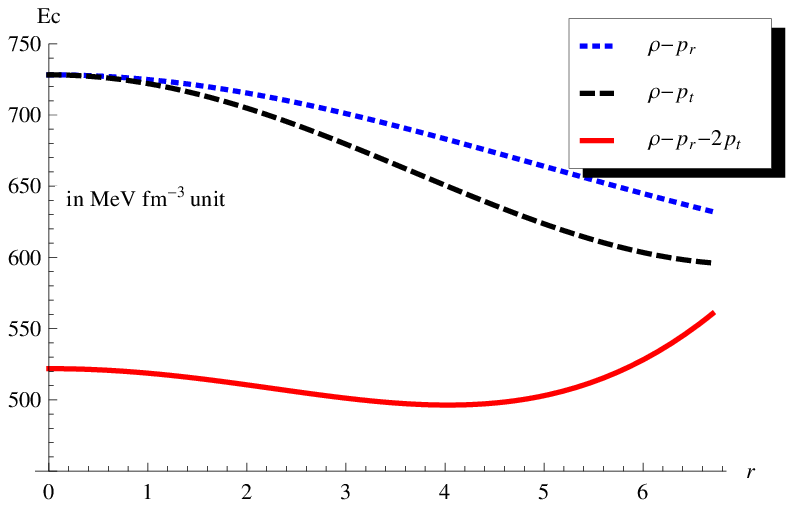}
      \caption{Weak, Null and Strong energy conditions are plotted against the radial distance $r$ inside the stellar interior by employing the same values of the constants as mentioned in Fig.~(\ref{rho}).}
   \label{ec}
\end{figure}

\section{Mass-radius relation}

The mass function in our model is obtained from (\ref{mass}) as
\begin{equation}
m(r)=\frac{br^3+15\left(r-\frac{\arctan
[\sqrt{b}r]}{\sqrt{b}}\right)}{6a^2} \label{mass1}
\end{equation}
The profile of the mass function is shown in Fig.~(\ref{mass}). Since at $r\rightarrow 0$ we have $m(r)\rightarrow 0$, it implies that the mass function is free from any central singularity.

Buchdahl\cite{buchdahl59} obtained an upper bound on the mass to radius ratio i.e., compactness $u$  of a relativistic star of a compact star such that $M/r < 4/9$.
In our model, we have
\begin{equation}
u(r)=\frac{br^3+15\left(r-\frac{\arctan
[\sqrt{b}r]}{\sqrt{b}}\right)}{6a^2 r}. \label{ur}
\end{equation}
The compactness $u$ for different compact star models are given in Table~2. The table shows that compactness of configurations are wihin the Buchdahl limit\cite{buchdahl59}.

We have also determined the surface redshift using the formula
 \[1+z_s=\left(1-\frac{2m}{R}\right)^{-\frac{1}{2}},\]
 which for our model turns out to be
 \begin{equation}
 z_s=\left(1-\frac{bR^3+15\left(r-\frac{\arctan
[\sqrt{b}R]}{\sqrt{b}}\right)}{3a^2 R}\right)^{-\frac{1}{2}}-1.
 \end{equation}
The values of the surface redshift parameter for different stellar configurations are given in Table~ 1. For an isotropic star, in the absence of a cosmological constant, Buchdahl\cite{buchdahl59} and Straumann \cite{stra} have shown that $z_s \leq 2$. B\"{o}hmer and Harko\cite{bh} showed that for an anisotropic star, in the presence of a cosmological constant, the surface redshift can take a much higher value $z_s \leq 5$. The restriction was subsequently modified by Ivanov \cite{ivanov} who showed that the maximum permissible value could be as high as $z_s = 5.211$. In our case, we have $z_s \leq 1$ for different compact star models developed in this paper.

\begin{table}
\centering
\begin{minipage}{140mm}
\caption{Values of the constants $a$, $b$, $H$ and $K$ for different compact star models.}
\begin{tabular}{@{}lrrrrrr@{}}
\hline
Compact Star & $R~$(km) & a & b & H & K & $M~$($M_{\odot}$) \\ \hline
Her X - 1   &6.7&1.3997&0.009&0.294&0.00012910&0.789  \\
RX J 1856-37 &6.006&1.6859&0.024&0.21 & 0.00020280& 0.904  \\
PSRJ 1614-2230 &11.3664&2.0952& 0.0138 &0.29   &0.000022416 & 1.86\\
SAX J1808.4-3658 & 7.07 &2.1015   & 0.036   & 0.29   & 0.000149572&1.157     \\ \hline
\end{tabular}
\end{minipage}
\end{table}

\begin{table}
\centering
\begin{minipage}{140mm}
\caption{Values of physical parameters for different compact star models. }
\begin{tabular}{@{}lrrrrrr@{}}
\hline
Compact Star & central density & surface density & central pressure & $u$ & $z_s$\\
& gm~cm$^{-3}$ & gm~cm$^{-3}$&dyne~cm$^{-2}$\\ \hline
Her X - 1   &$1.4805\times10^{15}$&$1.1255\times10^{15}$&$1.6535\times10^{35}$&0.1737&0.2379  \\
RX J 1856-37 &$2.7215\times10^{15}$&$1.6691\times10^{15}$&$3.2088\times10^{35}$&0.2219&0.3409   \\
PSRJ 1614-2230 &$1.0131\times 10^{15}$&$0.4722\times 10^{15}$&$2.0698\times10^{35}$& 0.2413 &0.3905\\
SAX J1808.4-3658 &$2.6271\times10^{15}$ &$1.2199\times 10^{15}$& $5.3784\times10^{35}$  & 0.2414  & 0.3906     \\ \hline
\end{tabular}
\end{minipage}
\end{table}

\section{Discussions}
In this paper we have presented a new model of a compact star in isotropic coordinates which is free from central singularity with the exterior being the vacuum Schwarzschild spacetime. To solve the Einstein field equations we have employed the modified Chaplygin equation of state which is inspired by the current observation of the expanding universe and its connection to the existence of dark energy. By considering the observed radius of the compact star Her X-1 as an input parameter (we have assumed $R=6.7~$km), we have analyzed physical viability of our model. We note that the metric coefficients are free from any singularity. The variations of $\rho$, $\frac{p_r+2p_t}{\rho}$ ,$p_r,~p_t$ are plotted in Fig.~(\ref{rho}), (\ref{diag}), (\ref{pr}) and (\ref{pt}), respectively which clearly show that matter density, radial and transverse pressure are positive inside the stellar interior and they are monotonically decreasing functions of radial coordinate. At the boundary of the star the matter density and transverse pressure are non-negative and the radial pressure vanishes as expected. From the plot of $\frac{p_r+2p_t}{\rho}$ (Fig.~(\ref{diag})) we note that it is non-zero and monotonically decreasing from the center towards the stellar surface. The anisotropic factor $\Delta=p_t-p_r$ is plotted against $r$ in Fig.~(\ref{delta}). Since $\Delta > 0$, the anisotropic factor is repulsive in nature in our model which is a desirable feature of a compact star\cite{gm}. Moreover, at the center of the star $\Delta$ vanishes which is also an essential feature of a realistic star. In order to investigate the relevance of our model in the study of compact stars, we have considered a number of compact stars, namely Her X-1, RX J 1856-37, PSRJ 1614-2230 and SAX J1808.4-3658 and showed that for the estimated radii the star's masses determined from our model are very close to the observed masses (see Ref.~\cite{maurya,tj}). This leads naturally to the proposition that the solution obtained in this paper can be used as a viable model for describing ultra-compact stars.

\end{document}